\documentclass[fleqn]{jfm}

\usepackage{epsfig}
\usepackage{amssymb}
\usepackage{amsmath}

\ifCUPmtlplainloaded \else
  \checkfont{eurm10}
  \iffontfound
    \IfFileExists{upmath.sty}
      {\typeout{^^JFound AMS Euler Roman fonts on the system,
                   using the 'upmath' package.^^J}%
       \usepackage{upmath}}
      {\typeout{^^JFound AMS Euler Roman fonts on the system, but you
                   dont seem to have the}%
       \typeout{'upmath' package installed. JFM.cls can take advantage
                 of these fonts,^^Jif you use 'upmath' package.^^J}%
      }
  \else
  \fi
\fi

\ifCUPmtlplainloaded \else
  \checkfont{msam10}
  \iffontfound
    \IfFileExists{amssymb.sty}
      {\typeout{^^JFound AMS Symbol fonts on the system, using the
                'amssymb' package.^^J}%
       \usepackage{amssymb}%
         \let\leq=\leqslant
         
      }{}
  \fi
\fi

\ifCUPmtlplainloaded \else
  \IfFileExists{amsbsy.sty}
    {\typeout{^^JFound the 'amsbsy' package on the system, using it.^^J}%
     \usepackage{amsbsy}}
    {}
\fi

\newsavebox{\astrutbox}
\sbox{\astrutbox}{\rule[-5pt]{0pt}{20pt}}

\newcommand\etal{\mbox{\textit{et al.}}}

\newcounter{saveeqn}
\renewcommand{\thesection}{\arabic{section}}

\renewcommand{\vec}[1]{\mbox{\boldmath $ #1$}}

\newcommand{\f}{\frac}

\title[Inertial Convection]{Inertial Convection in Rotating Fluid Spheres}

\author[F. H. Busse and R. Simitev]%
{F.\ns H.\ns B\ls U\ls S\ls S\ls E%
  \thanks{E-mail: Busse@uni-bayreuth.de}\ns
\and  R.\ns S\ls I\ls M\ls I\ls T\ls E\ls V\ns}

\affiliation{Institute of Physics, University of Bayreuth, D-95440
  Bayreuth, Germany} 

\pubyear{2003}
\volume{??}
\pagerange{??}
\date{?? and in revised form ??}
\begin{document}

\maketitle

\begin{abstract}
The onset of convection in the form of inertial  waves in a rotating  fluid
sphere is studied through a perturbation analysis in extension of earlier work
by \cite{Zhang94}. Explicit expressions for the dependence of the Rayleigh number
on the azimuthal wavenumber are derived and new results  for
the case of a nearly thermally insulating boundary are obtained.
\end{abstract}

\section{Introduction}
Convection in the form of slightly modified inertial waves is a well
known phenomenon in geophysical fluid dynamics. The analysis of the
onset of convection in a horizontal fluid layer heated from below and
rotating about a vertical axis was first done by Chandrasekhar more
than 50 years ago. For an account of this early work we refer to his
famous monograph (Chandrasekhar, 1961). He found that convection sets in at
high rotation rates in the form of modified inertial waves when the
Prandtl number is less than about 0.6 depending on the boundary
conditions. Another important case in which convection in the form of
modified inertial waves occurs is the rotating fluid sphere heated
from within and subject to a spherically symmetric gravity field. The
transition from convection in the form of columns aligned with the
axis of rotation to inertial convection in the form of equatorially
attached modes has been demonstrated by \cite{ZB87}. In a
later series of papers Zhang (1993, 1994, 1995) developed an analytical theory for
the critical parameter values for the onset of convection based on a
perturbation approach. The buoyancy term and viscous dissipation are
introduced in the equation of motion as small perturbations of inviscid
inertial waves and the balance of the two terms is the used for the
determination of the critical value of the Rayleigh number. In this
paper we extend this approach to case of a spherical boundary of low
thermal conductivity on the one hand and to an alternate method of
analysis on the other hand which will allow us to obtain explicit
expressions for the dependence of the Rayleigh number on the azimuthal
wave number.

\section{Mathematical Formulation of the Problem}

We consider a homogeneously heated, self-gravitating fluid sphere
rotating with the constant angular velocity $\Omega$ about an axis
fixed in space. A static state thus exists with the temperature
distribution $T_S = T_0 - \beta r_0^2 r^2 /2$ and the gravity field 
given by $\vec g = - \gamma r_0\vec r$ where $\vec r$ is the
position vector with respect to the center of the sphere and $r$ is
its length measured in fractions of the radius $r_0$ of the sphere. In
addition to the length $r_0$, 
the time $r_0^2 / \nu$ and  the temperature $\nu^2 / \gamma \alpha r_0^4$
are used as scales for the dimensionless description of the problem
where $\nu$ denotes the kinematic viscosity of the fluid and $\kappa$ is
its thermal diffusivity. The density is assumed to be constant except
in the gravity term where its temperature dependence given by $\alpha
\equiv ( d \varrho/dT)/\varrho =$ const. is taken into account.
The basic equations of motion and the heat equation for the deviation
$\Theta$ from the static temperature distribution are thus given by
\begin{subequations}
\label{100}
\begin{align}
\label{101}
&\partial_t \vec{u} + \tau \vec k \times
\vec u - \nabla \pi = \Theta \vec r + \nabla^2 \vec u, \\
\label{102}
&\nabla \cdot \vec u = 0, \\
\label{103}
&R\vec r \cdot \vec u + \nabla^2 \Theta - P\partial_t\Theta = 0,
\end{align}
\end{subequations}
where the Rayleigh number $R$, the Coriolis parameter $\tau$
and  the Prandtl number $P$ are defined by  
\begin{gather}
\label{110}
R = \frac{\alpha \gamma \beta r_0^6}{\nu \kappa} , \enspace \tau =
\frac{2 \Omega r_0^2}{\nu} , \enspace P = \frac{\nu}{\kappa}. 
\end{gather}
We have neglected the nonlinear terms  $\vec u \cdot \nabla \vec u$
and $\vec u \cdot \nabla \Theta$ in equations \eqref{100} since we
restrict the attention to the problem of the onset of convection 
in the form of small disturbances. In the limit of high $\tau$ the
right hand side of equation \eqref{101} 
can be neglected and the equation for inertial waves is obtained. For
the description of inertial wave solutions $\vec u_0$ we use 
the general representation in terms of poloidal and toroidal
components for the solenoidal field $\vec u_0$,  
\begin{gather}
\label{120}
\vec u_0 = \nabla \times ( \nabla v \times \vec r) + \nabla w \times
\vec r .
\end{gather}
By multiplying the (curl)$^2$ and the curl of the inertial wave equation by
$\vec r$ we obtain two equations for $v$ and $w$,  
\begin{subequations}
\label{130}
\begin{align}
\label{131}
& [ \partial_t {\cal L}_2 - \tau \partial_{\varphi} ] \nabla^2 v -
\tau {\cal Q} w =  0, \\
\label{132}
& [ \partial_t {\cal L}_2 - \tau \partial_{\varphi} ] w + \tau {\cal Q}v= 0 , 
\end{align}
\end{subequations}
where $\partial_t$ and $\partial_{\varphi}$ denote the partial
derivatives with respect to time $t$ and with respect to the  angle
$\varphi$ of a spherical system of coordinates $r, \theta, \varphi$
and where the operators ${\cal L}_2$ and ${\cal Q}$ are defined by  
\begin{subequations}
\label{140}
\begin{align}
\label{141}
&{\cal L}_2 \equiv - r^2 \nabla^2 + \partial_r ( r^2 \partial_r), \\
\label{142}
&{\cal Q} \equiv r \cos \theta \nabla^2 - ({\cal L}_2 + r \partial_r ) ( \cos \theta
\partial_r - r^{-1} \sin \theta \partial_{\theta}).
\end{align}
\end{subequations}
General solutions in explicit form for inertial waves in rotating
spheres have recently been obtained by \cite{Zhangetal01}. Here only
solutions of equations \eqref{130} for which $v$ is symmetric with respect to
the equatorial plane and does not possess a zero in its
$\theta$-dependence are of interest since only those are connected
with the preferred modes for onset of convection (Zhang, 1994). These modes are
given by 
\begin{gather}
\label{150}
\hspace*{-1.5cm}v_0 = P_m^m( \cos \theta ) \exp \{ im \varphi + i
\omega \tau t\} f (r), \quad 
w_0 = P_{m+1}^m ( \cos \theta ) \exp \{ i m \varphi + i \omega \tau t \} g (r) ,
\end{gather}
with
\begin{subequations}
\label{160}
\begin{align}
\label{161}
&\hspace*{-1.5cm} 
f ( r) = r^m - r^{m+2}, \quad 
g (r) =r^{m+1} \frac{2im(m+2)}{(2m+1)(\omega_0 (m^2+3m+2)-m)}
, \\
\label{162}
&\hspace*{-1.5cm}
\omega_0 = \frac{1 }{m+2} ( 1 \pm(1+m(m+2)(2m+3)^{-1})^{\frac{1}{2}} ).
\end{align}
\end{subequations}
Before considering the full problem \eqref{100} we have to specify the
boundary conditions. We shall assume a stress-free boundary with
either a fixed temperature (case A) or a thermally insulating boundary
(case B),
\begin{equation}
\label{180}
\vec r \cdot \vec u =  \vec r \cdot \nabla (\vec r \times \vec u) /r^2 = 0\quad  \mbox{   and } 
\begin{cases}
\Theta = 0  & \text{case A} \\
\partial_r\Theta = 0  & \text{case B} \end{cases} \quad \mbox{ at } \enspace r=1.
\end{equation}

Following \cite{Zhang94} we use a perturbation approach for solving
equations \eqref{100}, 
\begin{equation}
\label{190}
\vec u =\vec u_0 + \vec u_1 + ... , \quad  \omega = \omega_0 + \omega_1 + ...
\end{equation}
The perturbation $\vec u_1$ consists of two parts, $\vec u_1 = \vec
u_i + \vec u_b$ where $\vec u_i$ denotes the perturbation of the
interior flow, while $\vec u_b$ is the Ekman boundary flow which is
required since $\vec u_0$ satisfies  the first of conditions \eqref{180}, but
not the second one. 

After the ansatz \eqref{190} has been inserted into equations
\eqref{101} and \eqref{102} we obtain the solvability condition for
the equation \eqref{101} for $\vec u_1$ by multiplying it with  $\vec
u_0^*$ and averaging it over the fluid sphere, 
\begin{equation}
\label{200}
i\omega_1\langle|\vec u_0|^2\rangle = \langle\Theta\vec r \cdot \vec
u_0^*\rangle + \langle \vec u_0^* \cdot \nabla^2 (\vec u_0 + \vec u_b)\rangle,
\end{equation}
where the brackets $\langle...\rangle$ indicate the average over the
fluid sphere and the $*$ indicates the complex conjugate. The
evaluation of the second term on the right hand side of \eqref{200} yields 
\begin{equation}
\label{210}
\hspace*{-2cm}
\langle \vec u_0^* \cdot \nabla^2 (\vec u_0 + \vec u_b)\rangle = \langle (\nabla \times \vec u_0^*)\cdot(\nabla \times \vec u_b))\rangle + \frac{3}{4\pi}\oint [\vec u_0^*\cdot \nabla \vec u_b
- \vec u_0^*\cdot (\vec r\cdot \nabla) \vec u_b]d^2S,
\end{equation}
since $\nabla^2 \vec u_0$ vanishes (Zhang, 1994). Since $\vec u_b$ is of the
order $\tau^{-1/2}$ and vanishes outside a boundary layer of thickness
$\tau^{-1/2}$ only the term involving a radial 
derivative of $\vec u_b$ makes a contribution of the order one on the
right hand side of equation \eqref{210}. This term can easily be
evaluated because of the condition $\vec r \cdot \nabla \vec r \times
(\vec u_0 + \vec u_b)/r^2 = 0$ at the surface of the sphere. Using
expressions \eqref{150} and \eqref{161} we thus obtain
\begin{gather}
\label{220}
\hspace*{-1.5cm}
\langle \vec u_0^* \cdot \nabla^2 \vec u_b\rangle =
\frac{3}{2}\int_{-1}^{1}|P_m^m|^2 d( \cos\theta)\; m(m+1)(2m+1)\left[4 +
(m+2)\frac{(2m+1)}{2m+3}\right]  \nonumber \\
\hspace*{3.2cm}\cdot \left|\frac{2(m+1)^2-2}{(2m+1)(\omega_0(m+1)(m+2)-m)}\right|^2, 
\end{gather}
where the relationship
\begin{equation}
\label{230}
\int_{-1}^{1}|P_m^{m+1}|^2d cos\theta =
\frac{(2m+1)^2}{2m+3}\int_{-1}^{1}|P_m^m|^2d \cos\theta  
\end{equation}
has been used.

\section{Explicit Expressions in the Limit $P\tau \ll 1$}

The equation \eqref{103} for $\Theta$ can most easily be solved in the
limit of vanishing $\tau P \omega_0$.  In this limit we obtain for $\Theta$,
\begin{equation}
\label{240}
\Theta = P_m^m( \cos \theta ) \exp \{ im \varphi + i \omega \tau t \} h (r),
\end{equation}
with
\begin{gather}
\label{250}
\hspace*{-2cm}
h(r) = m(m+1)R\left(\frac{r^{m+4}}{(m+5)(m+4)-(m+1)m} -
  \frac{r^{m+2}}{(m+3)(m+2)-(m+1)m} - c r^m\right),
\end{gather}
where the coefficient $c$ is given by 
\begin{equation}
\label{260}
\hspace*{-1.5cm}
c = 
\begin{cases}
 \displaystyle\frac{1}{(m+5)(m+4)-(m+1)m} -
 \displaystyle\frac{1}{(m+3)(m+2)-(m+1)m}   & \text{case A,} \\ [10pt]
 \displaystyle\frac{(m+4)/m}{(m+5)(m+4)-(m+1)m} - \displaystyle\frac{(m+2)/m}{(m+3)(m+2)-(m+1)m}  & \text{case B.} \end{cases} 
\end{equation}
Since $\Theta$ is real $\omega_1$ must vanish according to the
solvability condition \eqref{200} and we  obtain for $R$ the final result
\begin{gather}
\hspace*{-1.5cm}
R_\pm = \left(\frac{m^2(m+2)^3}{(2m+3)[(m+1)(1\pm\sqrt{(m^2
      +4m+3)/(2m+3)} - m]^2} + 2m+1\right) \nonumber \\
\label{270}
\hspace*{4cm}
\cdot (2m+9)(2m+7)(2m+5)^2(2m+3)^2/b,
\end{gather}
where the two possibilities of the sign originate from the two
possibilities of the sign in the expression \eqref{162} for $\omega_0$. The
coefficient $b$ assumes the values 
\begin{equation}
\label{280}
b = 
\begin{cases}
 (m+1)m(10m+27)   & \text{case A,} \\[2pt]
 (m+1)(14m^2+59m+63)  & \text{case B.}
\end{cases} 
\end{equation}
Obviously the lowest value of $R$ is reached for $m=1$ and the value
$R_+$ for convection waves traveling in the retrograde direction is
always lower than the value $R_-$ for the prograde 
waves. Expression \eqref{270} is also of interest, however, in the case of spherical
fluid shells when the ($m=1$)-mode is affected most strongly by the
presence of the inner boundary. Convection modes corresponding to
higher values of $m$ may then  become preferred at onset since their
$r$-dependence decays more rapidly with distance from the outer boundary
according to relationships \eqref{160}.    

\section{ Solution of the Heat Equation in the General Case}

\begin{figure}
\epsfig{file=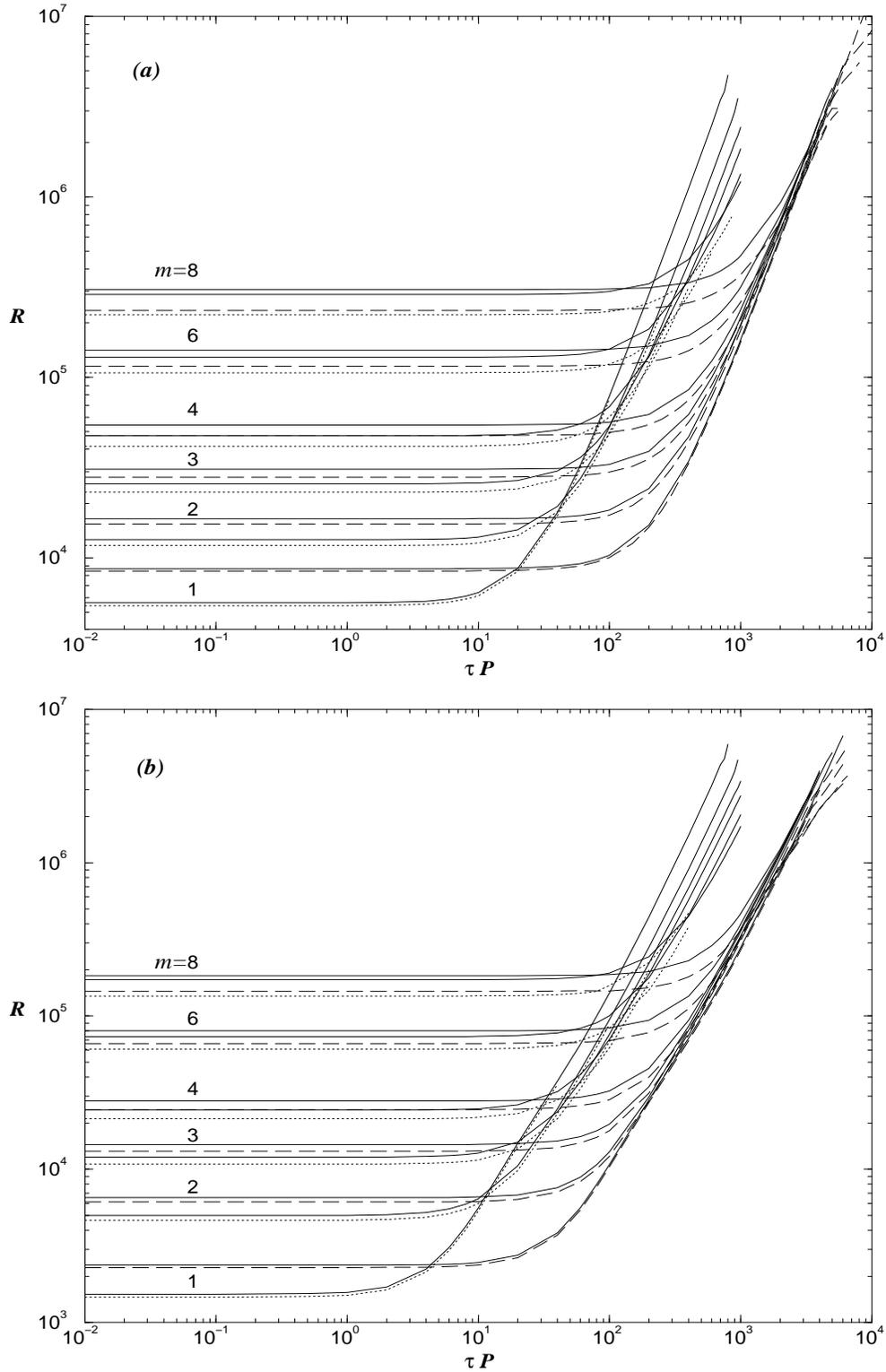,width=13cm,angle=0,clip=}
  \caption{ The Rayleigh number $R$ as a function of $\tau P$ for
    $m=1,2,3,4,6$ and $8$. Results based
    on explicit expressions such as \eqref{330} in the case of $m=1$
    (solid lines) are 
    shown in comparison with the results obtained with a Galerkin
    numerical scheme (dotted lines for retrograde mode, dashed lines
    for prograde mode).
    (\textit{a}) Case A, fixed temperature boundary conditions, 
    (\textit{b}) Case B, insulating thermal boundary conditions.}\label{f1}
\end{figure}
For the solution of equation \eqref{103} in the general case it is convenient
to use the Green's function method. The Green's function $G(r,a)$ is
obtained as solution of the equation 
\begin{equation}
\label{290}
 \left[\partial_r r^2 \partial_r + \big(-i \omega_0 \tau P\; r^2 - m(m+1)\big)\right] G(r,a)  = \delta(r-a),
\end{equation}
which can be solved in terms of the spherical Bessel functions $j_m(\mu
r)$ and $y_m(\mu r)$,
\begin{equation}
\label{300}
 G(r,a) = \begin{cases}
 G_1(r,a) = A_1 j_m(\mu r) \qquad & \text{for $0 \leq r < a$}, \\[2pt]
 G_2(r,a) = A j_m(\mu r)+ B y_m(\mu r)  \qquad & \text{for $a<  r \leq 1$, }
\end{cases}
\end{equation}
where
\refstepcounter{equation}
$$
\hspace*{-1.65cm}
\mu \equiv \sqrt{- i \omega_0 \tau P}, \quad
A_1= \mu\left(y_m(\mu a) - j_m(\mu a) \f{j_m(\mu)}{y_m(\mu)}
  \right), 
\eqno{(\theequation{\mathit{a},\mathit{b}})}  
$$
$$
\hspace*{-3.5cm}
  A= - \mu j_m(\mu a) \f{y_m(\mu)}{j_m(\mu)}, \quad 
  B= \mu j_m(\mu a). 
\eqno{(\theequation{\mathit{c},\mathit{d}})}  
$$
A solution of the equation \eqref{103} can be obtained in the form
\begin{gather}
\label{310}
\hspace*{-1.5cm}
h(r) = -\int_0^1 G(r,a)m(m+1)(a^m-a^{m+2}) a^2 d a =  \\
\hspace*{-1.5cm}
-\int_0^r G_2(r,a) m(m+1)(a^m-a^{m+2})
a^2 d a -\int_r^1 G_1(r,a) m(m+1)(a^m-a^{m+2}) a^2 d a . \nonumber
\end{gather}
Evaluations of these integrals for $m=1$ yield the expressions
\begin{equation}
\label{320}
\hspace*{-1.5cm}
h(r) = 
\begin{cases}
\displaystyle\frac{2R}{(\omega_0 \tau P)^2}\left( r(\mu^2 + 10) - \mu^2r^3 -
  \displaystyle\frac{10\big(\mu r\cos(\mu r) - \sin(\mu r)\big)}{r^2\big(\mu \cos\mu -
    \sin\mu \big)}\right)   & \text{case A,} \\ [15pt]
\displaystyle\frac{2R}{(\omega_0 \tau P)^2}\left( r(\mu^2 + 10) - \mu^2r^3 -
  \displaystyle\frac{(\mu^2-10)\big(\mu r \cos(\mu r) - \sin(\mu r)\big)}{r^2\big(2\mu \cos\mu -
    (2-\mu^2)\sin\mu\big)}\right)  & \text{case B.} 
\end{cases} 
\end{equation}
Slightly more complex expressions are obtained for $m
>1$. Expressions \eqref{320} can now be used to calculate $R$ and $\omega_1$
on the basis of equation \eqref{200}. In the case $m=1$ we obtain 
\begin{gather}
\label{330}
\hspace*{-1.5cm}
R = 21(\omega_0 \tau P)^2 \left(1 + \frac{9}{5(6\omega_0 - 1)^2}\right) \\ 
\hspace*{1.8cm}
\cdot  
\begin{cases}
\left[2-1050\mu^{-4}-\Re  \left \{\displaystyle\frac{350\mu^{-2}\sin\mu}{\mu
 \cos\mu-\sin\mu}\right\}\right]^{-1}  & \text{case A,} \\ [10pt]
 \left[ 9+525\mu^{-4}-\Re \left\{\displaystyle\frac{(7\mu^2-70+175\mu^{-2}) \sin\mu}{2\mu \cos\mu+(\mu^2-2) \sin\mu}\right\}\right]^{-1} & \text{case B,} \\
\end{cases} \nonumber
\end{gather}
where $\Re\{\}$ indicates the real part of the term enclosed by
$\{\}$. Expressions \eqref{330} have been plotted together with the
expressions obtained for higher values of $m$ in figures \ref{f1}a and \ref{f1}b
for the cases A and B, respectively.   We also show numerical values
by broken lines which have been obtained through the use of a modified
version of the Galerkin method of \cite{Ardes97}. Because the
numerical computations have been done for the finite value $10^5$ of
$\tau$ the results differ slightly from those of the analytical
theory.  Since there are two values of
$\omega_0$ for each $m$, two functions $R(\tau P)$ have been plotted for
each $m$. For values $\tau P$ of the order unity or lower, expressions
\eqref{270} are well approached and the retrograde mode corresponding to the
positive sign in \eqref{162}  yields always the lower
value of $R$. But it looses 
its preference to the prograde mode corresponding to the negative sign  in
\eqref{162} as $\tau P$ becomes of the order $10$ or larger depending on
the particular value of $m$. This transition can be understood on the
basis of the increasing difference in phase between $\Theta$ and $u_r$
with increasing $\tau P$. While the mode with the largest absolute
value of $\omega_0$ is preferred as long as $\Theta$ and $u_r$ are in
phase, the mode with the minimum absolute value of $\omega$ becomes
preferred as the phase difference increases since the latter is
detrimental to the work done by the buoyancy force.  The frequency
perturbation $\omega_1$
usually makes only a small contribution to $\omega$ which tends to
decrease the absolute value of $\omega$. 
\begin{figure}
\begin{center}
\hspace*{-0.5cm}\epsfig{file=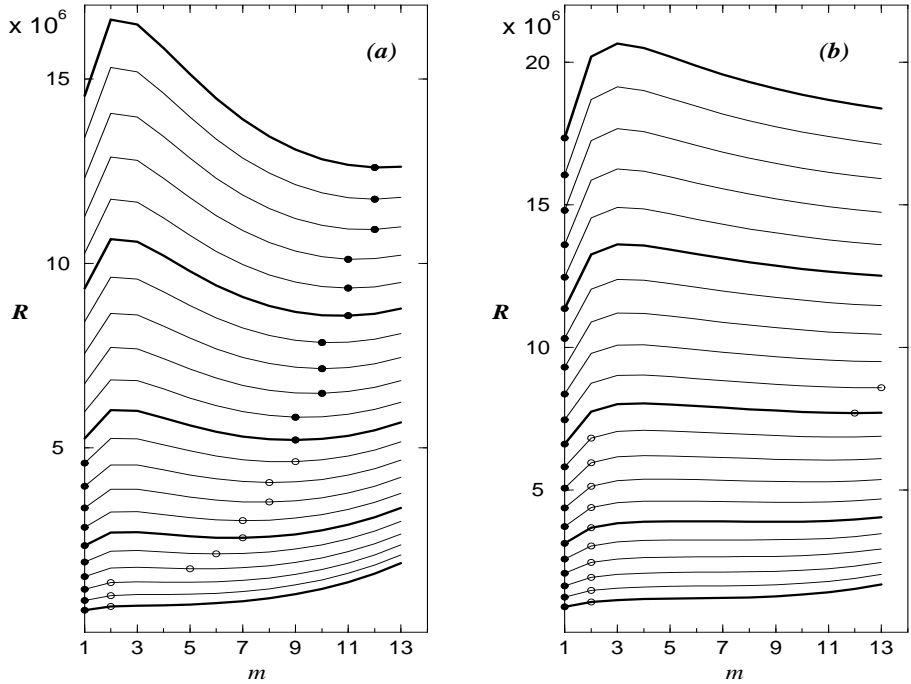,height=12cm,width=9cm,angle=-90,clip=}
\end{center}
  \caption{The Rayleigh number $R$ as a function of $m$ for
    $2 \cdot 10^3 \leq \tau P \leq 10^4$ (from bottom to top). The lines are equidistant with a
    step of $\Delta(\tau P) = 400$. The filled circles indicate the
    preferred values of $m$. The empty
    circles correspond to the preferred value of $m$ in the case when
    $m=1$ is not included in the competition.
    (\textit{a}) Case A, fixed temperature boundary conditions, 
    (\textit{b}) Case B, insulating thermal boundary conditions.}
\label{f2}
\end{figure}

For very large values of $\tau P$ the Rayleigh number  $R$ increases
in proportion to 
$(\tau P)^2$ for fixed $m$. In spite of this strong increase $\Theta$
remains of the order $\tau P$ on the right hand side of equation
\eqref{101}. The perturbation approach thus continues to be valid for
$\tau \longrightarrow \infty$ as long as $P \ll 1$ can be assumed. For
any fixed low Prandtl
number, however,   the onset of convection in the form of  prograde
inertial modes will 
be replaced with increasing $\tau$ at some point by the onset in the
form of columnar 
convection because the latter obeys an approximate asymptotic relationship for $R$
of the  form $(\tau P)^{4/3}$ (see, for example, \cite{Bu70}). This
second transition depends on the value of $P$  and will occur at
higher values of $\tau$ and $R$ for lower values of $P$. There is
little chance that inertial convection occurs in the Earth's core,
for instance, since $P$ is of the order $0.03$ while the usual estimate
for $\tau$ is $10^{15}$.

\section{Discussion}

Since the curves $R(\tau P, m)$ intersect at values of the order
$10^3$ of $\tau P$ in figures \ref{f1}a and \ref{f1}b, a different way of plotting
the results has been adopted in figure \ref{f2}. Here the preferred value of
$m$ has been indicated by a filled circle in the case of the prograde inertial
mode. The results of  figure \ref{f2}a agree well with those of figure 4 of
Zhang (1994) even though only an approximate method had been used for
the determination of the Rayleigh number. Zhang neglected the
$(m=1)$-mode and thus arrived at a different criterion for the
preferred mode. His preferred values  of $m$ are indicated by open
circles in figure 2a. The $(m=1)$-mode could indeed be suppressed by
the presence of an inner concentric spherical boundary. A rough
estimate indicates that inertial convection with the azimuthal
wavenumber $m$ will be affected significantly when the radius $\eta$ 
of the inner boundary exceeds a value of the order $(1 -
m^{-1})$. Unfortunately an analytical theory of inertial waves in 
rotating spherical fluid shells does not exist and it is thus not
possible to extend the analysis of this paper to the case when an
inner boundary is present. For a numerical study of inertial
convection in rotating spherical fluid shells and its finite
amplitude properties we refer to the paper of \cite{Simitev03}.

The two transitions between modes of different types mentioned in the
preceding section illuminate some of the puzzling findings of
\cite{ZB87} and \cite{Ardes97}. The transition labeled I in 
figure 17 of \cite{ZB87} can now be clearly identified with
the transition from  retrograde to prograde inertial convection. The
main result of our analysis is that this transition depends primarily
on the parameter combination $\tau P$ with only a minor dependence on
the  wavenumber $m$. The second transition from inertial to columnar
convection can not be pinned down as well because of the lack of a
sufficiently accurate analytical theory for thermal Rossby waves in
the low Prandtl number regime. According to the numerical results of
\cite{Ardes97} (see their figures 4 and 5) there exists a
broad transition range involving perhaps several transitions where the
onset of convection occurs in the form of multi-cellular modes. An
illumination  of this regime should be the goal of future research. \\[5pt]

\noindent{\large \bf Acknowledgment:} The research reported in this paper has
been performed in parts by the authors during their stay at the Woods
Hole Summer Program in Geophysical Fluid Dynamics 2002. The research
has also been supported by the {\it Deutsche Forschungsgemeinschaft}
under Grant Bu589/10-2.

\end{document}